# Shape Coexistence in Hot Rotating $^{100}$Nb

**MAMTA AGGARWAL**

Department of Physics, University of Mumbai, kalina Campus, Mumbai 400 098.

**Email: mamta.a4@gmail.com**

**Abstract** Temperature and angular momentum induced shape changes in the well deformed $^{100}$Nb have been investigated within the theoretical framework of Statistical theory combined with triaxially deformed Nilson potential and Strutinsky prescription. Two shape coexistence, one in the ground state of $^{104}$Nb between oblate and triaxial shapes and another one between oblate and rarely seen prolate non-collective shapes in excited hot rotating $^{100}$Nb at the mid spin values around 14-16h are reported for the first time. The level density parameter indicates the influence of the shell effects and changes drastically at the shape transition. The band crossing is observed at the sharp shape transition.

**Keywords:** Statistical theory; shape transition, A= 80-100, level density parameter, shape coexistence

## 1. INTRODUCTION

The hot rotating compound nucleus which is a many-body system with a complex internal structure, is treated in the framework of Statistical Model [1-3] with temperature, spin, isospin and deformation degrees of freedom within a mean field approximation with excitation energy and angular momentum as the input parameters. With increasing excitation energy, the density of quantum mechanical states increases rapidly and the nucleus shifts from discreteness to the quasi-continuum and continuum where the statistical concepts especially the nuclear level density (NLD) [4-9], which is the number of excited levels around an excitation energy, are crucial for the prediction of various nuclear phenomena, astrophysics [10] and nuclear technology. The excitation due to the temperature and rotation alter the nuclear structure significantly. The evolution of shapes [11-13] and phase transitions in such excited hot and rotating nuclei can be studied experimentally by the measurement of the GDR

Aggarwal, M

gamma rays [14]. Sometimes two shape phases appear to coexist with similar energies leading to the phenomenon of shape coexistence [15-16] studied in our earlier works [2,17-19]. The shape phase transitions in excited nuclei also impact the level density and the particle emission spectra shown in our recent work [20,21] and has become a subject of current scientific interest experimentally and theoretically.

Here we present our results on the shape evolution and coexistence in the odd Z (=41) Nb isotopes with A=80-100. This mass region A= 80-100 [22,23] is known to provide exotic nuclear structural phenomena often characterized by the shape coexistence. Since the large parts of N and Z are distributed in *fpg* shells, thus their level density is high and there is an interplay of the single-particle and collective motion. Also the intruder of the $1g_{9/2}$ orbitals located just above the N = 40 sub-shell closure plays an important role in the shape coexistence hence propose an ideal region to study shape evolution with spin, excitation and isospin. We also investigate the influence of the shape transitions on the level density parameter.

## 2. THEORETICAL FORMALISM

To evaluate deformation and shape of the excited nucleus we calculate excitation energy E* and entropy S of the hot rotating nuclear system using the statistical theory of hot rotating nuclei [1-3] for fixed temperature T and angular momentum M (given as an input) as a function of Nilson deformation parameter $\beta$ and $\gamma$. The excitation energy E* is incorporated with the ground state energy calculated using triaxially deformed Nilson Strutinsky method [17] and then the free Energy (F = E-TS) [24] is minimized with respect to deformation parameters ($\beta$, $\gamma$) at T and M.

$$F(Z,N,T,M,\beta,\gamma) = \left[ E_{LDM}(Z,N) + \delta E_{Shell}(\beta,\gamma) + E_{def}(Z,N,\beta,\gamma) + E^*(T,M,\beta,\gamma) \right] - T\, S(T,M,\beta,\gamma) \quad (1)$$

where $\delta E_{Shell}$ is the shell correction and $E_{def}$ is the deformation energy due to coulomb and surface effects. $E_{LDM}$(Z,N) is the macroscopic energy computed using Liquid drop mass formula. The excitation energy $E^*$(T,M,$\beta$,$\gamma$) is obtained as

$$E^*(T, M) = E(T, M) - E(0,0) \quad (2)$$

where E(0, 0) is the ground state energy. The rotational energy is given by

$$E_{rot}(M) = E(M,T) - E(0,T) \quad (3)$$

The level density parameter 'a' is computed as

$$a = S^2/ 4 E^* \quad (4)$$

and the Inverse level density parameter is obtained as K=A/a.

Free energy F minima are searched for various β (0 to 0.4 in steps of 0.01) and γ (from -180° (oblate) to -120° (prolate) and -180°< γ < -120° (triaxial)) to trace the nuclear shapes and equilibrium deformations. (The readers may refer to ref. [1-2] for the detailed theoretical formalism)

## 3. RESULTS AND DISCUSSION

We compute ground state deformation and shapes of the Nb isotopes (Z=41) with N = 40-66, using the triaxially deformed Nilson potential and the Strutinsky's prescription which is adequately described in our earlier works [17]. The energy minima are traced for all the nuclei as a function of deformation parameters β and γ which give the equilibrium deformation and shape of each nucleus. Fig. 1 shows the plot of ground state energy E = ($E_{LDM}$(Z,N) + $\delta E_{Shell}$(β,γ)+

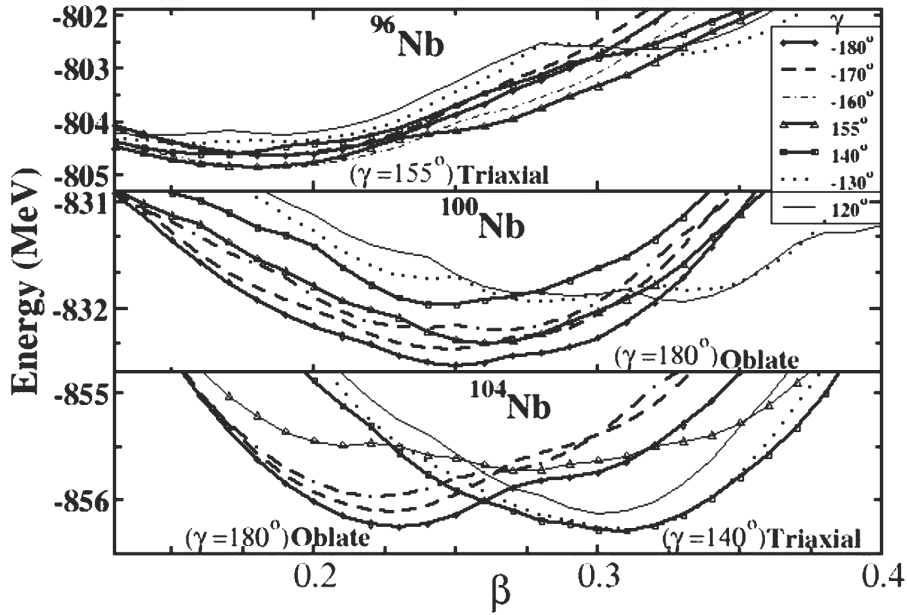

**Figure 1:** Energy E minima vs. β for various γ values for 96,100,104Nb. Shape coexistence at 104Nb is evident.

Aggarwal, M

$E_{def}(Z,N,\beta,\gamma))$ vs. $\beta$ for various $\gamma$ for $^{96,100,104}$Nb (since shapes of A= 98 and 102 are same as 100, we choose to study A = 96, 100, 104 with different shapes) where we find rapid shape transitions. E minima moves from the triaxial shape ($\gamma$ = -155°) at $^{96}$Nb, to oblate ($\gamma$ =-180°) at $^{100}$Nb with a well defined strong minima, to the coexisting oblate ($\gamma$ = -180°) and triaxial ($\gamma$ = -140°) shapes at $^{104}$Nb with large but different deformations and similar energy which identifies the phenomena of the shape coexistence in the ground state of $^{104}$Nb. Values of $\beta$ (= 0.17 (triaxial), 0.25 (oblate minima) with 0.33 (prolate minima at energy difference 606 KeV) and 0.24 (oblate minima) with 0.31 (triaxial minima)) of $^{96}$Nb, $^{100}$Nb and $^{104}$Nb respectively are in reasonable match with $\beta$ (= 0.18, 0.38 and 0.38) values with prolate shape of Ref. [25].

Fig. 2 (a) shows the deformation of $^{80-102}$Nb isotopes in ground as well as in excited states where we have included temperature degree of freedom. Ground state (GS) deformation (at T = 0) varying between 0.1-0.27 starts decreasing as the temperature increases and becomes zero at T=1.5 MeV. The dominant shape phase in this region is found to be triaxial with few oblate shapes. The inclusion of triaxial shapes [26] in our calculation makes them more meaningful especially in this region which is expected to have dominant triaxial deformation space. Deformation shows a minima at N=50 (A=91) although the deformation is not zero as is expected from a shell closure in this odd-even nucleus in the highly deformed region.

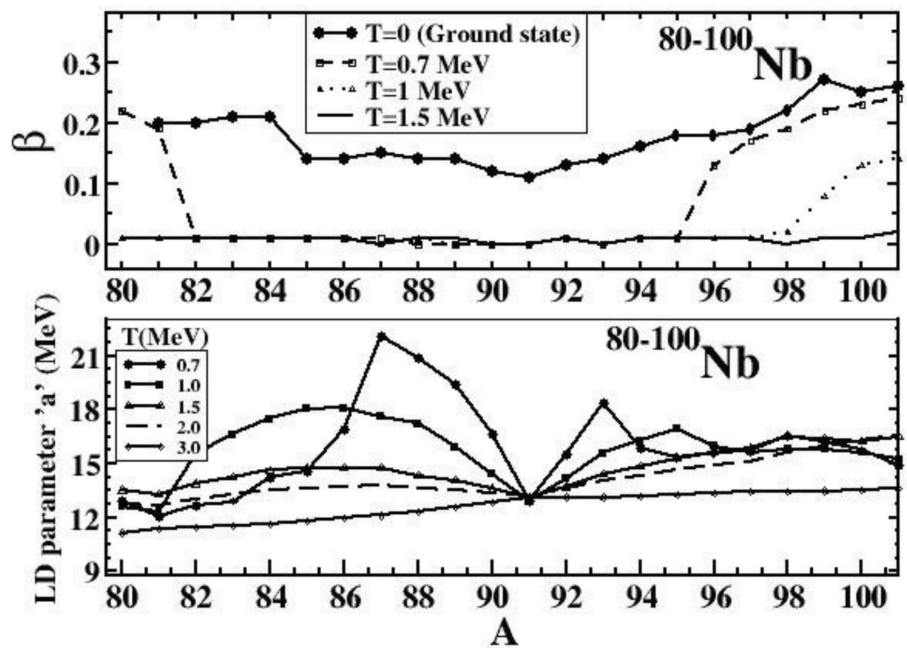

**Figure 2:** (a) $\beta$ vs. A and (b) level density (LD) parameter vs. A for various T=0, 0.7, 1.0, 1.5, 2.0 MeV showing the ground state(GS) and excited state of 80-100Nb isotopes.

Shell effects are evident in Fig. 2(b) where we have plotted level density (LD) parameter vs. A for various T = 0.7, 1.0, 1.5, 2.0 MeV. The LD parameter 'a' is a minima at shell closure and maxima in mid shell region at low temperature T=0.7 MeV. With increasing T, shell effects melt away and 'a' varies almost smoothly with slight increasing value with A.

Nuclear structure is strongly impacted as soon as we incorporate the collective and non-collective rotational degrees of freedom. Fig. 3 shows β (Fig. 3(a)) and γ (Fig. 3(b)) vs. angular momentum for various temperature values T=0.7, 1.0, 1.5, 2.0 MeV and M = 0.5 to 60.5 $\hbar$. The deformation is high even at a high temperature T=2 MeV and further increases with increasing M and reaches up to a value of 0.2. A close inspection of shapes in Fig. 3(b) reveals that at low temperature T = 0.7 MeV where the states are nearly yrast

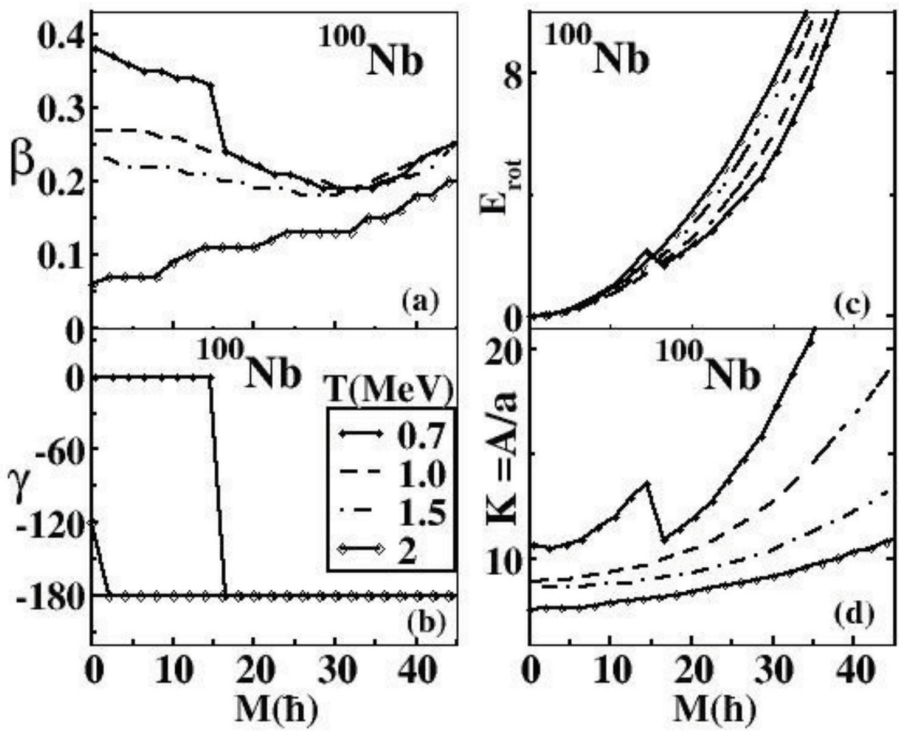

**Figure 3:** Variation of (a) β (b) γ, (c) Erot, and (d) K=A/a vs. M($\hbar$) = 0.5-45.5$\hbar$ is shown for various temperature values T=0.7, 1.0, 1.5, 2.0 MeV for 100Nb.

Aggarwal, M

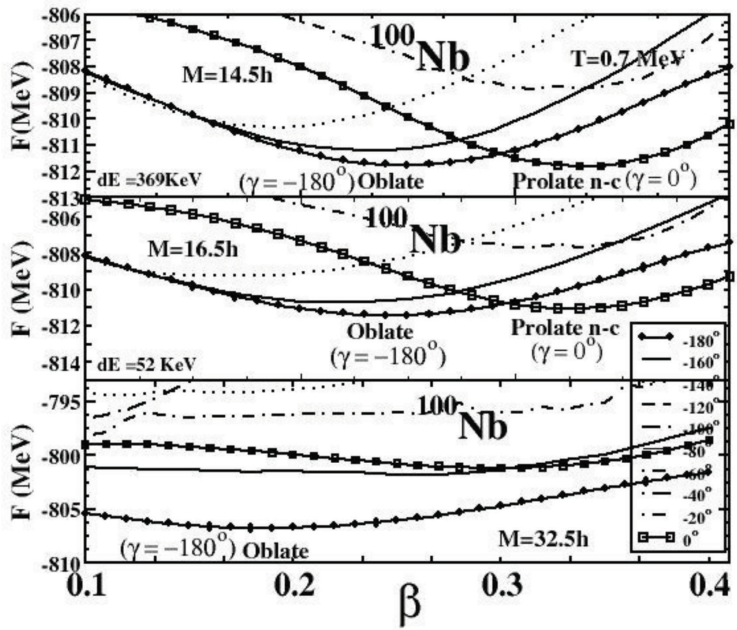

**Figure 4:** Free energy (F) minima vs. β for various γ for M= (a) 14.5, (b) 16.5 and (c) 32.5ℏ at T=0.7 MeV. Shape coexistence between Oblate (γ =−180o ) and triaxial (γ = −140o ) is seen at 16.5ℏ.

and the effects of rotation are dominant, we find a rarely seen shape phase of prolate non-collective, first anticipated by Goodman [27] and then observed in our earlier works [1-2,28], which diminishes as T and M increases. At this sharp shape transition, band crossing is observed which is evident in the plot of rotational energy vs. M. (in Fig. 3(c)). With increasing T, the band crossing effect diminishes and the rotational energy $E_{rot}$ varies gradually with angular momentum at all temperatures.

Fig. 3(d) shows the influence the sharp shape transition on level density parameter of a nucleus where we have plotted the inverse level density parameter K=A/a vs. M. Normally K increases as M increases because the part of the excitation energy is spent in rotation due to which the level density decreases and K increases. However, at the sharp shape transition at $M = 16.5\,\hbar$, we note a sharp drop in the value of K which in turn enhances the nuclear level density and hence the particle emission probability shown in our recent works [20,21] which has shown good agreement with recent measurements [29-30] on nuclear level density and emission spectra for medium and heavy mass region nuclei $^{119}$Sb and $^{185}$Re. Our observation [20] of enhancement of level density and drop in inverse level density parameter associated with the deformation and shape changes has provided important inputs to understand various predictions of experimental works [29-31]. However, the experimental and other theoretical data for the Nb isotopes investigated in this region is awaited.

The phenomenon of shape coexistence is observed in excited hot and rotating nucleus $^{100}$Nb not reported so far in any other work as far to our knowledge. In Fig. 4, we plot free energy F minima vs. β for various γ for M=14.5, 16.5 and 32.5 $\hbar$. We find that the F minima moves from prolate at M= 14.5 $\hbar$ (with oblate shape nearly coexisting with energy difference of 369 KeV) to an oblate F minima (slightly deeper than prolate) at M=16.5 $\hbar$ coexisting with prolate with merely an energy difference of 52 KeV which identifies a shape coexistence where the shape phases of prolate and oblate non-collective appear to coexist at similar energy but have very different deformations. At higher spin M=32.5 $\hbar$, F minima goes to the expected usual shape phase of oblate non-collective with a well defined single minima.

## CONCLUSION

Ground state and highly excited hot and rotating states of $^{80-100}$Nb isotopes are studied within a microscopic approach. Odd - Z Nb isotopes are found to have predominantly triaxial shapes and few oblate shapes with high deformation ranging between 0.1-0.35. Shape coexistence between oblate and triaxial shapes in ground state $^{104}$Nb is observed. The level density parameter shows the influence of shell effects. A sharp shape transition from prolate to oblate non-collective for hot rotating $^{100}$Nb leads to the effects of band crossing and a sharp drop in inverse level density parameter, which slowly fade away with the increasing temperature showing the structural effects disappearing as temperature increases. While undergoing the shape transition from prolate to oblate non-collective, we observe a shape coexistence with oblate and prolate non-collective shapes at mid spin value in $^{100}$Nb.

## ACKNOWLEDGMENT

We thank the financial support by SERB, DST, Govt. of India, under the women scientist scheme WOS-A. Support from Dr. G. Saxena is gratefully acknowledged.

Aggarwal, M